\documentclass[mathleft
]{an}
\usepackage{graphicx}
\usepackage{times}
\begin{document}

\Pagespan{1009}{1012}
\Yearpublication{2012}%
\Yearsubmission{2012}%
\Month{??}%
\Volume{333}%
\Issue{10}%
 \DOI{10.1002/asna.201211789}%

\title{On the surface structure of sunspots}

\author{M. Franz\inst{1}\fnmsep\thanks{Corresponding author:
  \email{morten@kis.uni-freiburg.de}\newline}
}
\titlerunning{The structure of sunspots}
\authorrunning{M. Franz}
\institute{
Kiepenheuer Institut f\"ur Sonnenphysik, Sch\"oneckstra\ss e 6, 
D-79104 Freiburg, Germany}
\received{2012 Aug 24}
\accepted{2012 Oct 16}
\publonline{2012 Dec 3}


\abstract{A precise knowledge of the surface structure of sunspots is essential to construct adequate input models for helioseismic inversion tools. 
We summarize our recent findings about the velocity and magnetic field in and around sunspots using HINODE observation. To this end we quantize the horizontal and vertical component of the penumbral velocity field at different levels of precision and study the moat flow around sunspot. Furthermore, we find that a significant amount of the penumbral magnetic fields return below the surface within the penumbra. Finally, we explain why the related opposite polarity signals remain hidden in magnetograms constructed from measurements with limited spectral resolution.}

\maketitle

\section{Introduction}
Sunspots are one of the most prominent manifestations of solar magnetic fields \cite{Hale:1908}. These spots consist of a central dark region, called the umbra, surrounded by a semi-dark ring, i.e. the penumbra. The latter shows a rich structure of bright and dark radially elongated penumbral filaments, cf.~e.g.~\cite{Secchi:1875} and Movie 1. Sunspots occur, because a strong (up to 3.5~kG) magnetic field disturbs the convective energy transport from the solar interior \cite{Bray:1964}. As a result, the solar surface cools radiatively to temperatures of 4000~K and appears darker when compared to the quiet Sun, which is at a temperature of about 5800~K \cite{Solanki:2003}. Like the magnetic activity, the number of sunspots show an 11\footnote{Since the polarity of the magnetic field is reversed every 11 years it is actually a 22 year cycle until the same solar hemisphere shows the same polarity again.} year dependence too \cite{Schwabe:1844}. In the beginning of the cycle they appear around $\pm30^{\circ}$ latitude. As the cycle progresses, the latitude where they appear decreases, and they are located around the equator towards the end of the cycle \cite{Carrington:1858}. 

It is consensus that the origin of sunspots is a bundle of strong magnetic flux that rises through the convection zone, penetrating the surface of the sun and ascending into the solar atmosphere \cite{Parker:1955}. At the footpoints of this flux loop, active regions with binary polarities are located \cite{Zwaan:1987}. Sunspots often appear within these active regions, where the leading polarity forms larger, more stable and longer living spots \cite{Fan:1993}. Note that active regions have opposite polarity in different hemispheres and different solar cycles \cite{Hale:1925}. While the general mechanism that is causing sunspots is widely accepted, the details are still under debate.

The major obstacle in our understanding is the lack of knowledge about the processes in the solar interior. How exactly rises the flux bundle through the convection zone? Is it shredded when it penetrates the solar surface? What is the exact structure of the sunspot magnetic field below the surface; is it a single monolithic flux tube or is it rather a conglomerate of individual flux bundles? What is the interplay between plasma flows and magnetic fields? Is the moat flow around sunspots of the same origin as the Evershed flow in the penumbra? How deep are these flows, and do they reverse their direction in deeper layers?

Some of these questions have been tackled by helioseismology, which provide the rare possibility to study the solar interior. However, it was found that different techniques of data evaluation yield different, sometime contradictory, results about the subsurface structure of sunspots \cite{Gizon:2009}. Furthermore, the subsurface structure is deduced in an inversion process which relies on theoretical models of sunspots \cite{Moradi:2010}. The situation is complicated by the fact that the magnetic fields within sunspots was not taken into account properly in the past.

To infer the subsurface structure of sunspots using helioseismic tools, it is essential to construct adequate sunspot models that reproduce the surface features as precisely as possible. To this end we summarize the current knowledge of the surface structure of sunspots. We focus on the properties of the plasma flow in and around sunspots as well as on the configuration of the surface magnetic field.

\section{Evershed Flow}

\begin{figure*}
\includegraphics[width=\textwidth]{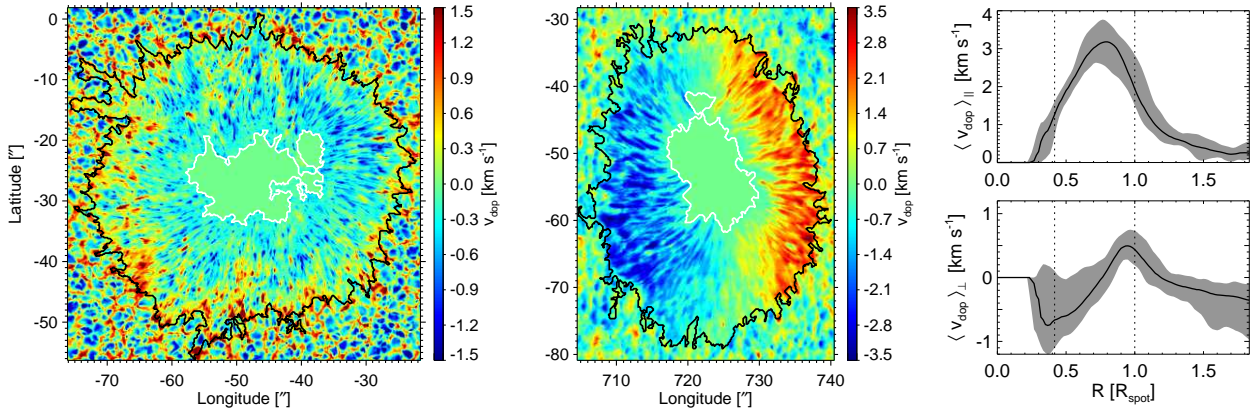}
\caption{Left: A Doppler map constructed from the shift of the line wing of Fe~I~630.15~nm. The white and black contours mark the inner and outer white light boundary of the spot. The umbral velocities have been artificially set to zero. Longitude and latitude are in solar disk coordinates. Since the sunspot is located at $\Theta = 3^{\circ}$ penumbral up- and down-flows are visible. Middle: Same as before, but $\Theta = 47^{\circ}$. Now the Evershed outflow, which is parallel to the surface, dominates the penumbral velocities. Right top: Average horizontal velocity along elliptical paths of increasing radii. The solid black line shows the average value of 17 sunspot samples at various heliocentric angles, while the maximal scatter of the individual measurements is indicated by the gray shaded region. The two vertical dashed lines indicate the inner and outer penumbral boundary. Right bottom: Same as before but average vertical velocity.}
\label{fig:franz_fig01}
\end{figure*}

Spectroscopic observations of the penumbra of sunspots yield not only a displacement, \cite{Evershed:1909}, but also an asymmetry of the solar spectral line, e.g. \cite{Holmes:1961}. Such observations can be explained by assuming a radially outwards directed flow of plasma that is present in a thin atmospheric layer in the deep photosphere, cf.~\cite{Maltby:1964}. This so-called Evershed flow stops abruptly at the white light boundary of sunspots, e.g. \cite{Schmidt:1992}. 
Evidence exists that a small fraction continues along the canopy into the chromosphere \cite{Rezaei:2006}, while the majority submerges in strong plasma downdrafts located at the outer penumbral boundary \cite{Franz:2009}.

Fig.~\ref{fig:franz_fig01} shows the penumbral plasma velocities of NOAA 10933 observed with the HINODE spectropolarimeter on 5$^{\rm{th}}$ (disk center) and 9$^{\rm{th}}$ (solar limb) of January 2007, cf.~\cite{Franz:2010}. The left panel demonstrates than up- and down-flows occur throughout the penumbra. Up-flows are elongated and appear predominately in the inner penumbra. Down-flows are roundish and cluster at the outer boundary. The plasma stream can reach speeds up to 5~km~s$^{-1}$ in individual down-flow channels, while the maximum up-flow velocity is $-$2~km~s$^{-1}$. Up- and down-flows in the inner and outer penumbra can be interpreted as the sources and the sinks of the Evershed flow.

The panel in the middle shows a sunspot at large heliocentric angles ($\Theta = 47^{\circ}$), where the horizontal component of the Evershed flow dominates. Besides the overall blue and redshift of the center and limb side penumbra, a filamentary structure with strong azimuthal variations is apparent. Noteworthy is the absence of Dopplershifts in the inner limb side penumbra, which allow an estimate of the inclination of the up-flow channels. If they have a zenith angle of $90^{\circ} - \Theta \approx 40^{\circ}$ they appear perpendicular to the line of sight and the Doppler signal vanishes. Remarkable are the patches of redshifts in the outer center side penumbra, e.g. at $\rm{(x;y)}=(708.$\arcsec$5;-56.$\arcsec$5)$ with v$_{\rm{dop}}=1$~km~s$^{-1}$. They indicate that some down-flow channels in the outer penumbra have an inclination of $90^{\circ} + \Theta \approx 135^{\circ}$ \cite{Franz:2011}. 

To quantify the vertical and horizontal velocities in the penumbra, we follow the procedure described in \cite{Schlichenmaier:2000}. If the penumbral flow field is assumed to be axially symmetric, then a sinusoidal fit to the datapoints extracted from an azimuthal path in the penumbra yields the average vertical (offset) and average horizontal (amplitude) line of sight velocity along that path. See Movie 2 for an illustration of this procedure. Fig.~\ref{fig:franz_fig01} shows the radial dependence of the average plasma velocity parallel (top right) and perpendicular (bottom right) to the solar surface. The thick solid line indicates the average values from 17 observations of the disk passage of NOAA 11147 in 2011 between January 22$^{\rm{nd}}$ and 25$^{\rm{nd}}$, while the gray shaded region mark the extrema of the individual measurements. Finally, both curves where deprojected using cos($\Theta$) for the horizontal and sin($\Theta$) for the vertical component.

Despite the large scatter of the individual measurements, we find that the plasma flow parallel to the surface increases in the inner penumbra, shows a maximum of 3~km~s$^{-1}$ at a radial distance $\rm{R} = 0.8~\rm{R}_{\rm{spot}}$, and drops to 2~km~s$^{-1}$ at the outer penumbral boundary. Note that this number is the average value of an elliptical path. However, since the outer edge of the sunspot is not elliptical, but rather a corrugated line, these numbers overestimate the real horizontal velocities in the immediate surrounding of the sunspot. The average vertical flow is negative $<$$\rm{v}_{\rm{dop}}$$>_{\perp} \approx -0.75$~km~s$^{-1}$ in the inner penumbra before it reverses its direction. It reaches $<$$\rm{v}_{\rm{dop}}$$>_{\perp} \approx 0.5$~km~s$^{-1}$ in the outer penumbra, before it becomes negative again in the quiet Sun. These negative values are expected because of the convective blueshift of the latter. The large scatter of the individual curves could be due to the center to limb variation of the convective blueshift, which was not taken into account in the individual measurements, as well as the lack of p-mode filtering of the datasets.


\section{Moat Flow}

\cite{Sheeley:1972} was one of the first to report on a radial outwards directed flow in the periphery of sunspots. This so-called moat flow develops after the formation of the spot, its velocity ranges from 0.5~km~s$^{-1}$ to 1~km~s$^{-1}$, and it extends to roughly twice the spot radius \cite{Brickhouse:1988}. Contrary to the Evershed flow, which is believed to be a surface phenomenon, helioseismic techniques provide evidence that the moat flow continues with speeds of 1~km~s$^{-1}$ for $3\cdot10^7$~m and seems to be present in depths of 2000~km \cite{Gizon:2000}.


Our results are shown in the top right panel of Fig.~\ref{fig:franz_fig01}. For $\rm{R}_{\rm{spot}} \le \rm{R} <1.2~\rm{R}_{\rm{spot}}$ the average horizontal velocity ($<$$\rm{v}_{\rm{dop}}$$>_{\parallel}$) drops rapidly from 2~km~s$^{-1}$ to 0.8~km~s$^{-1}$. These numbers are significantly higher than the previously reported values, and might be due to the fact that, at this distance, we still sample part of the penumbral Evershed flow in filaments extending into the moat region. The decrease of $<$$\rm{v}_{\rm{dop}}$$>_{\parallel}$ is less rapid for $1.2~\rm{R}_{\rm{spot}} \le \rm{R} < 1.7~\rm{R}_{\rm{spot}}$ and reaches 0.2~km~s$^{-1}$ at $\rm{R} \approx 1.7~\rm{R}_{\rm{spot}}$. For greater distances the results are not trustworthy anymore, since the fit does not yield a significant reduction of $\chi^2$.

\section{Magnetic field}
At moderate spatial resolution (2 arcsec) the magnetic field of a sunspot may be approximated by a flux tube \cite{Solanki:2003}. The field is strongest in the central umbra, where it is perpendicular to the surface. The field strength decreases with increasing radial distance from the center of the spot, and the zenith angle increases, i.e. the field becomes more inclined \cite{Borrero:2011}.  Due to lateral pressure balance and the stratification of the solar atmosphere, the diameter of the fluxtube growths with increasing height and forms the superpenumbral canopy \cite{Giovanelli:1980}. However, at high spatial resolution, the magnetic field inclination ($\gamma$) shows strong azimuthal variations, i.e. fluted \cite{Title:1993} or uncombed \cite{Solanki:1993} penumbra. Evidence exist that $\gamma$ is larger in the dark penumbral filaments when compared to the bright ones 
and there have been reports of penumbral magnetic fields returning into the Sun \cite{Westendorp:1997}.

The time evolution of the vertical magnetic field is shown by Movie 3. It shows NOAA 10933 observed with the narrow band filter imager (NFI) of HINODE on 2007 Janurary 5$^{\rm th}$ between 04.22 UT and 08.17 UT. The cadence of the dataset is two minutes and the spatial resolution is 0.\arcsec32. 
Even though the NFI suffers from problems of the data quality \cite{Ichimoto:2008}, and no calibration of the magnetic field strength is available \cite{Lamb:2010}, the movie is good enough for a qualitative study. 
It nicely illustrates the uncombed structure of the penumbral magnetic field. Since the spot is observed close to disk center, large value of Stokes V 
appear in regions where the field is more vertical (black). Since the field is more inclined -- maybe it is also weaker -- between these regions, the Stokes V signal is weak (grey) there. Note that only a very small region of the penumbra shows magnetic fields with opposite polarity (white). In the moat region, magnetic patches are visible running away from the spot. It is believed that these moving magnetic features remove flux from the spot and thus play an important role in the process of decay of sunspots.

Puzzling is the absence of opposite polarity patches in penumbral magnetograms. This is because a number of theoretical models \cite{Schlichenmaier:2002,Thomas:2002} predict the magnetic field to return below the surface within or just outside the penumbral boundary. The absence of opposite polarity has furthermore been taken as an argument against such penumbral models \cite{Spruit:2006}. However, using spectropolarimetric observation from HINODE we were able to identify a significant amount of magnetic fields returning below the solar surface within the body of the penumbra \cite{Franz:2012}:

\begin{figure*}
\includegraphics[width=0.98\textwidth]{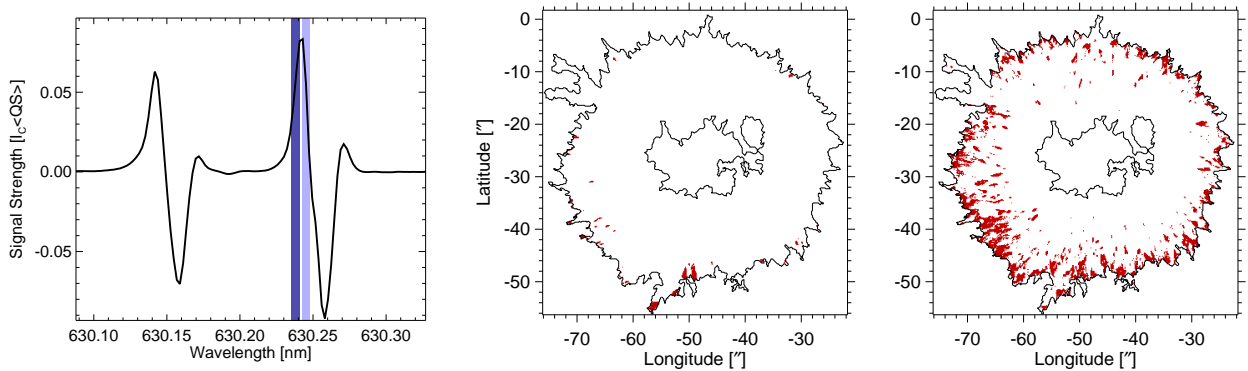}
\caption{Left: A three lobed Stokes V profile of penumbral down-flow channels as measured by the HINODE spectropolarimeter. The vertical strips (light and dark blue) represent the spectral regions in which some filter instruments measure the Stokes profiles in order to construct a classical magnetogram. Middle: Penumbral magnetic fields of opposite polarity (red) as derived from a classical magnetogram. Right: A magnetogram in which the third lobe of the asymmetric Stokes V profile is taken into account. Magnetic fields with "hidden" opposite polarity (red) are detectable in 40\% of all penumbral down-flows.}
\label{fig:franz_fig02}
\end{figure*}

To this end it is important characterize the asymmetry of penumbral Stokes V profiles as precisely as possible. The reason for that is illustrated in the left panel of Fig.~\ref{fig:franz_fig02}. The solid black line shows a typical Stokes V profile from a penumbral down-flow region with an additional lobe on the red side of a regular profile. In the most simple scenario such a three lobed Stokes V profile is the superposition of an antisymmetric and unshifted profile and another strongly red-shifted profile of opposite polarity. This third lobe is thus a signature of magnetic field of opposite polarity in the deep photospheric layers of penumbral down-flows.

Usually magnetograms are constructed from observation with filter instruments that measure a small spectral window slightly detuned from the line core. \cite{Langhans:2005}, for example, measures at $-$50~m\AA~with FWHM 72~m\AA~(cf.~light blue stripe in Fig.~\ref{fig:franz_fig02}) and HINODE NFI (Movie 3) measures at $-$120~m\AA~ with FWHM 60~m\AA~(dark blue stripe). Since the third lobe is located at least $+$180~m\AA~away from the line core, magnetic fields of opposite polarity remain hidden in such observation.

This is illustrated in the middle panel of Fig.~\ref{fig:franz_fig02} where magnetic fields of opposite polarity (indicated in red) are present in only 5\% of all penumbral down-flows. In the right panel we show all penumbral down-flows that contain Stokes V profiles with a third lobe. Now the amount of return flux in penumbral down-flows increases to 40\% and is only limited by the signal to noise ratio of the measurement.

\section{Discussion}

In this contribution we have summarized the properties of the velocity and magnetic field in and around sunspots.

A comparison of probability density functions, computed from maps of Doppler velocity, show that the penumbral flow pattern is significantly different from that of the quiet Sun \cite{Franz:2009}. Up- and down-flows appear throughout the penumbra and can be seen as the sources and sinks of the Evershed flow. Up-flows occur predominately at the inner boundary and have peak velocities of $-$2~km~s$^{-1}$. Down-flows cluster the outer edge of the sunspot and reach velocity amplitudes of 5~km~s$^{-1}$. On average the plasma flow parallel to the surface increases monotonously to 3~km~s$^{-1}$ at $\rm{R} \approx 0.8~\rm{R}_{\rm{spot}}$ before it drops to 2~km~s$^{-1}$ at the outer edge of the sunspot. The unusual high values of the moat flow -- $\rm{v}_{\rm{dop}} > 1$~km~s$^{-1}$ for $\rm{R}_{\rm{spot}} \le \rm{R} \le 1.2~\rm{R}_{\rm{spot}}$ -- are probably an artifact of our analysis method, since the jagged outer edge of the spot can not be sampled properly by an ellipse. 

The evolution of the sunspot magnetic field is studied using HINODE filtergrams. Similar to white light images, the penumbra shows a filamentary structure in circular polarization too. This is due to the strong azimuthal variation of the inclination of the penumbral magnetic field \cite{Title:1993}. The time series shows small magnetic patches running away from the spot. These so-called moving magnetic features are believed to play a key role in processes of decay of sunspots \cite{Kubo:2008}. Furthermore, we find a significant amount of magnetic fields returning below the surface in the outer penumbra. We demonstrate that the related opposite polarity signals remain hidden in classical magnetograms which are constructed from filter instruments with limited spectral resolution.

\acknowledgements
M. Franz wants to thank O. Steiner, R. Rezaei, W. Schmidt and R. Schlichenmaier for fruitful discussions and valuable comments on the manuscript. Part of this work was supported by the Deutsche Forschungsgemeinschaft, DFG project number Schl. 514/3-2. Hinode is a Japanese mission developed and launched by ISAS/JAXA, with NAOJ as domestic partner and NASA and STFC (UK) as international partners. It is operated by these agencies in cooperation with ESA and NSC (Norway).

\end{document}